\documentclass[a4paper,11pt]{article}
\usepackage{pos}

\newcommand{\be}{\begin{equation}}
\newcommand{\ee}{\end{equation}}

\title{Finite volume effects in the extended linear sigma model via low momentum cutoff}

\author*[a,b]{Győző Kovács}
\author[a]{Péter Kovács}
\author[c]{Pok Man Lo}
\author[c]{\linebreak Krzysztof Redlich}
\author[a]{György Wolf\vspace{0.5cm}}

\affiliation[a]{Institute for Particle and Nuclear Physics, Wigner Research Centre for Physics,\\
  Konkoly-Thege Miklós út 29–33, 1121 Budapest, Hungary}

\affiliation[b]{ELTE Eötvös Loránd University, Institute of Physics,\\
Pázmány Péter sétány 1/A, 1117 Budapest, Hungary}

\affiliation[c]{Institute of Theoretical Physics, University of Wroclaw, \\PL-50204 Wrocław, Poland}

\emailAdd{kovacs.gyozo@wigner.hu}

\abstract{Contrary to field theoretical calculations in the thermodynamic limit where the volume is assumed to be infinitely large, the heavy-ion collisions always carry the effects of finite size. A sufficiently small system size is expected to affect the thermodynamic quantities and the phase diagram of the strongly interacting matter. To study these effects one can take into account the finite spatial extent of the system within the framework of an effective model too, via the restriction of the momentum integrals using discretization or in a simplified case using a low momentum cutoff. We investigated the effects of the finite volume in a vector meson extended Polyakov quark-meson model and found a remarkable change in the thermodynamics and the phase transition, especially in the location of the critical endpoint. 
}

\FullConference{%
  FAIR next generation scientists - 7th Edition Workshop (FAIRness2022)\\
  23-27 May 2022\\
  Paralia (Pieria, Greece)
}

\begin{document}
\maketitle

\section{Extended linear sigma model}
The vector and axial vector meson Extended Polyakov Linear Sigma Model (ELSM) is an advanced quark-meson model, including four full meson nonets and $2+1$ flavor constituent quarks in the fermion sector. The ELSM was already used to study the thermodynamics and the phase diagram of the strong interaction in \cite{Kovacs:2016juc,Kovacs:2021kas}, where all the details and also the parameterization of the model can be found. 
The thermodynamics determined from the mean-field level grand potential 
\be \nonumber
 \Omega (T, \mu_q)= U_{Cl}+ \Omega_{\bar{q}q}^{(0)}(T,\mu_q) + U (\Phi ,\bar{\Phi}) 
\ee
built up from the classical potential, the fermionic one-loop contribution containing a momentum integration, which could be modified to include the finite volume effects, and the Polyakov loop potential. The field equations are given by minimizing the grand potential in the order parameters, $\phi_N$, $\phi_S$, $\Phi$, and $\bar \Phi$, that are the scalar-isoscalar meson condensates and the Polyakov loop variables, respectively.

\section{Finite volume effects on physical quantities and the phase diagram}

The finite volume effects are usually studied within effective field-theoretical models by taking into account the discretization in momentum space given by the finite spatial extent in the direct space. This method is used in many cases \cite{Palhares:2009tf,Tripolt:2013zfa} and also in the most recent studies \cite{Bernhardt:2021iql}. Although it does not include each aspect of the finite-sized physical system, it can be a good approximation to study the difference between infinite or large volume systems -- like usual theoretical models or the compact stars -- and small systems like the fireball in a heavy-ion collision. A further simplification can be made by using a low momentum cutoff instead of discretization, by making use of the observation that the zero momentum modes are the most relevant ones for the criticality of the system. A similar approximation was also studied in the HRG model and showed a good agreement with the direct finite volume calculations \cite{Karsch:2015zna}. In the present study, we used this method with a lower cut $\lambda=\pi/L$ in the integration in $\Omega_{\bar{q}q}^{(0)}(T,\mu_q)$ to implement the effects of the finite system size $L$.

\begin{figure}
	\centering
        \includegraphics[width=0.98\linewidth]{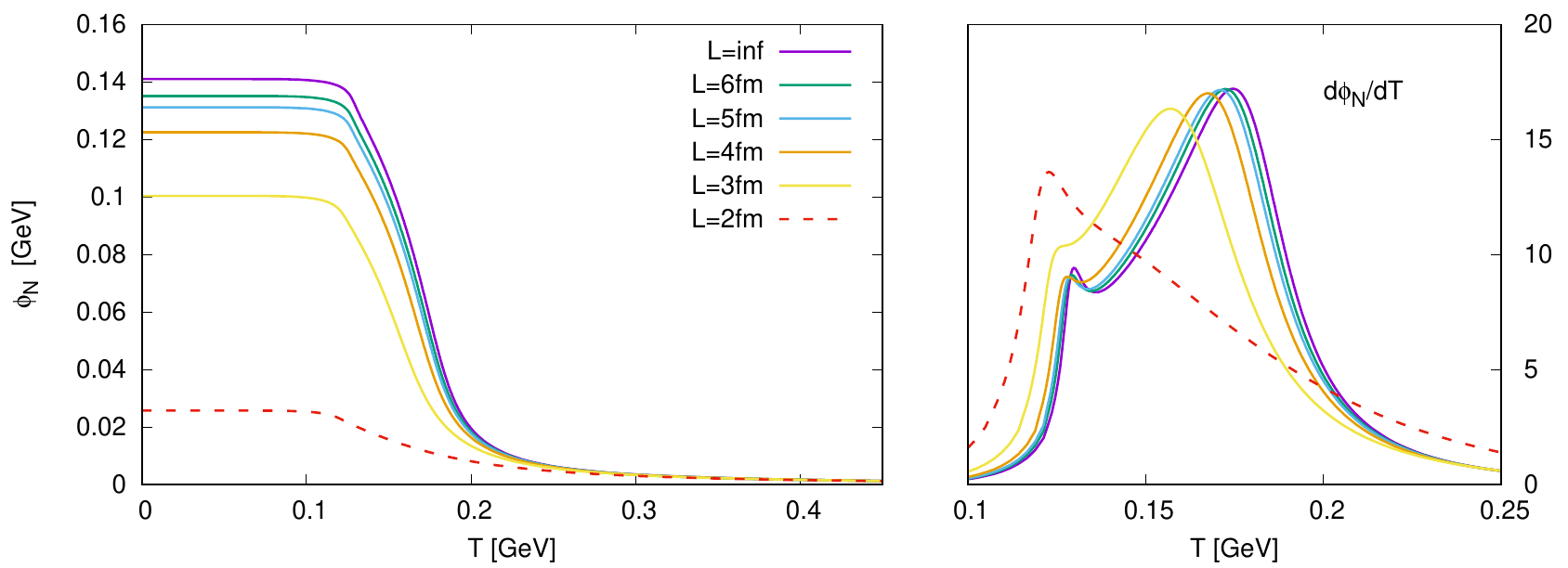}
		\caption{The temperature dependence of the non-strange meson condensate and its derivative for different system sizes. The dashed line signals a qualitatively different behavior.} \label{fig1}
\end{figure}
	
\begin{figure}
        \centering
	\includegraphics[width=0.49\linewidth]{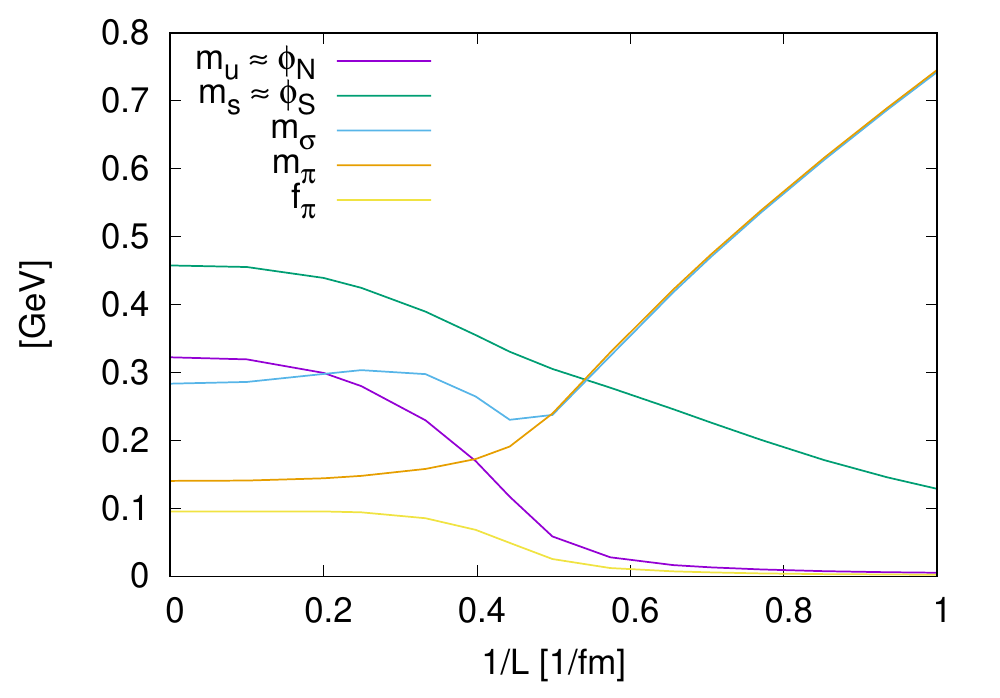}
	\includegraphics[width=0.49\linewidth]{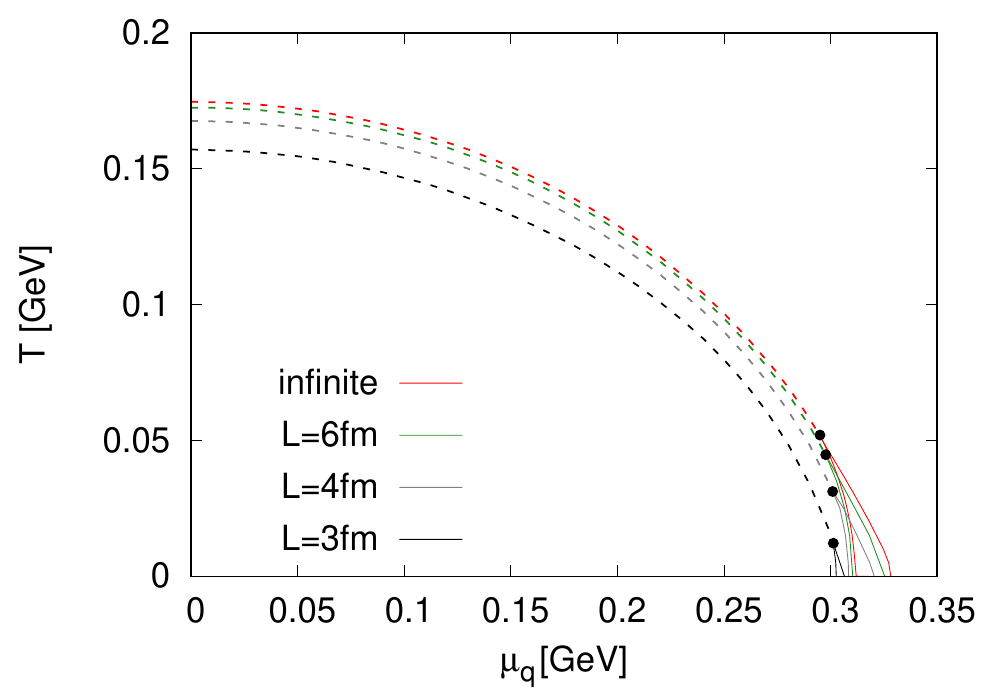}
		\caption{The modification of physical quantities at $T=0$, $\mu=0$ with $1/L$ (left panel). The finite size dependence of the phase diagram and the CEP (right panel). } \label{fig2}
\end{figure}

The solution of the modified field equations -- by the implementation of finite size  -- changes, compared to its infinite volume version,  and already the vacuum value of the meson condensates 
decreases with decreasing volume as shown in Fig.~\ref{fig1}. 
Physical quantities depending on the condensates directly are changing accordingly and they show a crossover-like behavior in $1/L$ that is presented in Fig.~\ref{fig2}. 
The $1/L$ dependence shows -- very similarly to the $T$ dependence -- that for larger values the system is in the chiral symmetric phase. We should also note, that below a certain size validity of models with a thermodynamic approach is also questionable.
	
We also investigated the phase diagram which is shown on the right panel of Fig.~\ref{fig2}. It was found that the transition temperature at $\mu_q=0$ decreases with the decreasing volume, while the transition becomes smoother. At the same time, the CEP moves to lower temperatures and slightly higher chemical potentials until it disappears around $L=2.5$ fm after which the criticality of the system is washed out.

As a continuation, we plan to implement also discretization in the momentum space and compare the results in the two scenarios and study other interesting quantities like baryon fluctuations that are often used in the investigation of the critical endpoint.

\vspace{.1cm}
\section{Acknowledgement}

The research was supported by the Hungarian National Research, Development and Innovation Fund under Project No. FK 131982. K.R. and P.M.L. acknowledge support from the Polish National Science Center (NCN) under Opus grant no. 2018/31/B/ST2/01663. K.Gy. thank the PHAROS COST Action (CA16214) for partial support.

\end{document}